\documentclass[a4paper,12pt]{article}
\usepackage{amssymb}
\usepackage{latexsym}
\usepackage[dvips]{graphicx}
\usepackage{cite}
\newcommand{\be}{\begin{equation}}
\newcommand{\ee}{\end{equation}}
\newcommand{\bea}{\begin{eqnarray}}
\newcommand{\nn}{\nonumber}
\newcommand{\eea}{\end{eqnarray}}

\begin{document}

\begin{titlepage}
\begin{flushright}
UB-ECM-PF-03/30
\end{flushright}
\begin{centering}
\vspace{.3in}
{\Large{\bf Effective mass of a radiating charged particle\\ in Einstein's universe}}
\\

\vspace{.5in} {\bf  Elias C.
Vagenas\footnote{evagenas@ecm.ub.es} }\\

\vspace{0.3in}

Departament d'Estructura i Constituents de la Mat\`{e}ria\\
and\\ CER for Astrophysics, Particle Physics and Cosmology\\
Universitat de Barcelona\\
Av. Diagonal 647\\ E-08028 Barcelona\\
Spain\\
\end{centering}

\vspace{0.7in}
\begin{abstract}
\par\noindent
The effective gravitational mass as well as the energy and momentum
distributions of a radiating charged particle in Einstein's universe are evaluated.
The M{\o}ller's energy-momentum complex is employed for this computation. The spacetime
under study is a generalization of Bonnor and Vaidya spacetime in the sense that the metric
is described in the cosmological background of Einstein's universe in lieu of
the flat background. Several spacetimes are limiting cases of the one considered here.
Particularly for the Reissner-Nordstr\"om black hole background,
our results are exactly the same with those derived by
Cohen and Gautreau using Whittaker's theorem and by Cohen and de Felice using Komar's mass.
Furthermore, the power output for the spacetime under consideration
is obtained.
\end{abstract}

\end{titlepage}
\newpage

\baselineskip=18pt
\section*{Introduction}
Energy-momentum localization has been one of the most interesting but also thorny problems
for the General Theory of Relativity. A plethora of different attempts to solve this problem have
led to inconclusive results till now. Energy-momentum complexes introduced first by Einstein
\cite{einstein}, were the foremost endeavor to solve this problem. After that a large number
of different expressions for the energy-momentum complexes were proposed \cite{others}.
A drawback of this attempt was that energy-momentum complexes had to be computed
in quasi-Cartesian coordinates. M{\o}ller \cite{moller} proposed a new expression
for an energy-momentum complex which could be utilized to any coordinate system.
However, the idea of the energy-momentum complex was severely criticized
for a number of reasons \cite{drawbacks}.
Considerable attempts to deal with this problematic issue are also
the quasi-local \cite{quasilocal} and the superenergetic quantities \cite{superenergy}
\footnote{There is some interest in employing the energy-momentum complexes in the
framework of teleparallel equivalent of General Relativity, i.e. teleparallel gravity \cite{teleparallel}.}.

\par\noindent
Virbhadra and collaborators enlivened anew the concept of energy-momentum complexes \cite{virbhadra}.
Since then, numerous works have been performed on evaluating the energy and momentum distributions
of several gravitational backgrounds using the energy-momentum complexes \cite{complexes}.
In support of the importance of the concept of energy-momentum complexes,
Chang, Nester and Chen \cite{nester} proved that every energy-momentum complex is associated with a
Hamiltonian boundary term. Thus, the energy-momentum complexes are quasi-local and acceptable.

\par\noindent
In this paper we evaluate the energy and momentum density distributions of a
radiating charged particle in Einstein's universe.
Additionally, the total energy and the power output are computed.
The total energy is actually the effective gravitational
mass whose gravitational field a neutral particle experiences.
The prescription that is used in the present analysis, is the one introduced by M{\o}ller.
The reasons for presenting here the M{\o}ller's description are:
(a) the argument that it is not restricted to quasi-Cartesian coordinates and (b) a
work of Lessner \cite{lessner} who argues that the M{\o}ller's energy-momentum complex is a powerful concept of energy
and momentum in General Theory of Relativity.

\par\noindent
The remainder of the paper is organized as follows.
In Section 1 we consider the concept of energy-momentum complexes in the framework
of General Theory of Relativity.
In Section 2 the M{\o}ller's energy-momentum complex is presented.
In Section 3 we give the metric that describes a radiating charged particle in Einstein's universe.
In Section 4 we utilize M{\o}ller's prescription and we calculate the energy and momentum density distributions
of the afore-mentioned spacetime. We also compute explicitly the effective gravitational mass and the
power output for the specific background. Furthermore, we evaluate the above-mentioned quantities for several
spacetimes which are limiting cases of the one presented in Section 3. We compare the results derived in the
present analysis with the ones that already exist in the literature. Finally,
Section 5 is devoted to a brief summary of results and concluding remarks.
\section{Energy-Momentum Complexes}
The conservation laws of energy and momentum  for an isolated (closed), i.e. no external force acting on the
system, physical system in the Special Theory of Relativity are expressed by a set of differential equations.
Defining $T^{\mu}_{\nu}$ as the symmetric energy-momentum tensor of matter and all non-gravitational fields the
conservation laws are given by
\be
T^{\mu}_{\nu,\, \mu} \equiv \frac{\partial T^{\mu}_{\nu}}{\partial
x^{\mu}}=0\ee where \be \rho=T^{t}_{t}\hspace{1cm}j^{i}=T^{i}_{t}\hspace{1cm}p_{i}=-T^{t}_{i}
\ee
are the energy density, the energy current density, the momentum density,
respectively, and Greek indices run over the spacetime
labels while Latin indices run over the spatial coordinate values.

\par\noindent
Making the transition from the Special to General Theory of Relativity one adopts a simplicity principle which is
called principle of minimal gravitational coupling. As a result of this, the conservation equation is now written
as \be T^{\mu}_{\nu;\, \mu} \equiv \frac{1}{\sqrt{-g}}\frac{\partial}{\partial
x^{\mu}}\left(\sqrt{-g}\,T^{\mu}_{\nu}\right)-\Gamma^{\kappa}_{\nu\lambda}T^{\lambda}_{\kappa}=0 \ee where $g$ is
the determinant of the metric tensor $g_{\mu\nu}(x)$. The conservation equation may also be written as \be
\frac{\partial}{\partial x^{\mu}}\left(\sqrt{-g}\,T^{\mu}_{\nu}\right)=\xi_{\nu}\ee where \be
\xi_{\nu}=\Gamma^{\kappa}_{\nu\lambda}T^{\lambda}_{\kappa}\ee is a non-tensorial object. For $\nu=t$ this means
that the matter energy is not a conserved quantity for the physical system\footnote{It is possible to restore the
conservation law by introducing a local inertial system for which at a specific spacetime point $\xi_{\nu}=0$ but
this equality by no means holds in general.}. From a physical point of view this lack of energy conservation can
be understood as the possibility of transforming matter energy into gravitational energy and vice versa. However,
this remains a problem and it is widely believed that in order to be solved one has to take into account the
gravitational energy \cite{others} .\par\noindent By a well-known procedure, the non-tensorial
object $\xi_{\nu}$ can be written  as \be \xi_{\nu}=-\frac{\partial}{\partial
x^{\mu}}\left(\sqrt{-g}\,\vartheta^{\mu}_{\nu}\right)\ee where $\vartheta^{\mu}_{\nu}$ are certain functions of
the metric tensor and its first order derivatives. Therefore, the energy-momentum tensor of matter $T^{\mu}_{\nu}$
is replaced by the expression \be \theta^{\mu}_{\nu}=\sqrt{-g}\left(T^{\mu}_{\nu}+\vartheta^{\mu}_{\nu}\right)\ee
which is called energy-momentum complex since it is a combination of the tensor $T^{\mu}_{\nu}$ and a pseudotensor
$\vartheta^{\mu}_{\nu}$ which describes the energy and  momentum of the gravitational field. The energy-momentum
complex satisfies a conservation law in the ordinary sense, i.e.
\be
\theta^{\mu}_{\nu,\, \mu}=0
\ee
and it can be
written as
\be
\theta^{\mu}_{\nu}=\chi^{\mu\lambda}_{\nu,\,\lambda}
\ee
where $\chi^{\mu\lambda}_{\nu}$ are called
superpotentials and are functions of the metric tensor and its first order derivatives.

\par\noindent
It is obvious that the energy-momentum complex is not
uniquely determined by the condition that is usual divergence is zero since it can always been added to the
energy-momentum complex a quantity with an identically vanishing divergence.

\section{M{\o}ller's Prescription}
The energy-momentum complex of M{\o}ller in a  four-dimensional background
is given as \cite{moller}
\be
\mathcal{J}^{\mu}_{\nu}=\frac{1}{8\pi}\xi^{\mu\lambda}_{\nu\,\, ,\, \lambda}
\label{mtheta}
\ee
where the M{\o}ller's superpotential $ \xi^{\mu\lambda}_{\nu}$ is of the form
\be
\xi^{\mu\lambda}_{\nu}=\sqrt{-g}
\left(\frac{\partial g_{\nu\sigma} }{\partial x^{\kappa} }-
\frac{\partial g_{\nu\kappa}}{\partial x^{\sigma}
}\right)g^{\mu\kappa}g^{\lambda\sigma}
\label{msuper}
\ee
with the antisymmetric property
\be
\xi^{\mu\lambda}_{\nu}=-\xi^{\lambda\mu}_{\nu}\hspace{1ex}.
\ee

\par\noindent
It is easily seen that the M{\o}ller's energy-momentum complex
satisfies the local conservation equation
\be
\frac{\partial \mathcal{J}^{\mu}_{\nu}}{\partial x^{\mu}}=0
\ee
where  $\mathcal{J}^{0}_{0}$ is the energy density and $\mathcal{J}^{0}_{i}$ are the momentum density components.

\par\noindent
Thus, the energy and momentum in M{\o}ller's
prescription for a four-dimensional background are given by
\be
P_{\mu}=\int\int\int
\mathcal{J}^{0}_{\mu}dx^{1}dx^{2}dx^{3}
\label{mmomentum}
\ee
and specifically the energy of the physical system in a
four-dimensional background is
\be
E=\int\int\int
\mathcal{J}^{0}_{0}dx^{1}dx^{2}dx^{3}\hspace{1ex}.
\label{menergy}
\ee

\par\noindent
It should be noted that the calculations are not anymore restricted
to quasi-Cartesian coordinates but they can be
utilized in any coordinate system.
\section{A radiating charged particle in Einstein's universe}
In 1970 Bonnor and Vaidya \cite{vaidya} presented a solution describing a radiating charged particle in
flat spacetime. The corresponding metric is of the form
\be
ds^{2} = 2dudr + \left(1-\frac{2M(u)}{r}+\frac{4\pi Q^{2}(u)}{r^{2}}\right)du^{2} -
r^{2}\left(d\theta^{2}+\sin^{2}\theta d\phi^{2}\right)
\label{bonnor}
\ee
where $u$ is the retarded null coordinate, i.e. $u=t-r$, and $M(u)$ and $Q(u)$ are respectively the mass
and charge of the particle. The mass function $M(u)$ is an arbitrary nonincreasing function of
the retarded null coordinate $u$.
The particle lives in a flat background and this is easily seen by letting
the radial coordinate go to infinity, i.e. $r\rightarrow\infty$.

\par\noindent
Patel and Akabari \cite{patel} realized that it would be more interesting to have the particle in a
cosmological background.
Therefore, they considered the space surrounding the radiating charged particle to be occupied by
a spherical symmetric matter distribution of nonzero density $\rho$ and pressure $p$.
Finally, they derived the following metric
\bea
ds^{2} &=& 2dudr + \left(1-\frac{2M(u)}{R}\cot\left(\frac{r}{R}\right)+
\frac{4\pi Q^{2}(u)}{R^{2}}\left[\cot^{2}\left(\frac{r}{R}\right) -1\right]\right)du^{2}\nn\\
&&- R^{2}\sin^{2}\left(\frac{r}{R}\right)\, \left(d\theta^{2}+\sin^{2}\theta\, d\phi^{2}\right)
\label{patel}
\eea
where $R$ is a constant and by setting the mass and the charge of the particle equal to zero,
namely $M(u)=Q(u)=0$, the metric (\ref{patel}) reduces to the metric of Einstein's universe
\be
ds^{2} = 2dudr + du^{2} -
R^{2}\sin^{2}\left(\frac{r}{R}\right)\, \left(d\theta^{2}+\sin^{2}\theta\, d\phi^{2}\right)
\hspace{1ex}.
\label{cosmo}
\ee
\par\noindent
It is noteworthy that several spacetimes are limiting cases of the
gravitational background under study.
\section{The effective mass}
The aim of this section is to evaluate the effective gravitational mass of the radiating charged particle
in Einstein's universe using the M{\o}ller's energy-momentum complex. We first have to evaluate
the superpotentials in the context of M{\o}ller's prescription.
There are twelve nonzero superpotentials
\bea
\xi^{2\,1}_{1}&=&-\xi^{1\,2}_{1}=-2\left[M(u)- \frac{4\pi Q^{2}(u)}{R}\cot\left(\frac{r}{R}\right)\right)]\sin\theta\nn\\
\xi^{3\, 1}_{3}&=&-\xi^{1\, 3}_{3}=-2R\sin\left(\frac{r}{R}\right)\cos\left(\frac{r}{R}\right)\sin\theta\nn\\
\xi^{4\, 1}_{4}&=&-\xi^{1\, 4}_{4}=-2R\sin\left(\frac{r}{R}\right)\cos\left(\frac{r}{R}\right)\sin\theta\nn\\
\xi^{2\, 3}_{3}&=&2\left(2M(u)\cos^{2}\left(\frac{r}{R}\right)-R\sin\left(\frac{r}{R}\right)\cos\left(\frac{r}{R}\right)\right.
\label{super}\\
 &&\left. -\frac{4\pi Q^{2}(u)}{R}\cot\left(\frac{r}{R}\right)\left[2\cot^{2}\left(\frac{r}{R}\right)-1\right]\right)\sin\theta\nn\\
\xi^{3\, 2}_{3}&=&-\xi^{2\, 3}_{3}\nn\\
\xi^{3\, 4}_{4}&=&-\xi^{3\, 4}_{4}=-2 \cos\theta\nn\\
\xi^{2\, 4}_{4}&=&-\xi^{4\, 2}_{4}=\xi^{2\, 3}_{3}\nn
\eea

\par\noindent
By substituting the M{\o}ller's superpotentials, as given by (\ref{super}), into equation (\ref{mtheta}),
one gets the energy density distribution
\be
\mathcal{J}^{0}_{0}=\frac{Q^{2}(u)}{R^{2}\sin^{2}\left(\frac{r}{R}\right)}\sin\theta
\label{energyden}
\ee
while the momentum density distributions take the form
\bea
\mathcal{J}^{0}_{1}&=&0\label{momden1}\\
\mathcal{J}^{0}_{2}&=&\frac{1}{4\pi}\sin\left(\frac{r}{R}\right)\cos\left(\frac{r}{R}\right)\cos\theta\label{momden2}\\
\mathcal{J}^{0}_{3}&=&0\label{momden3}\hspace{1ex}.
\eea

\par\noindent
Therefore, if we substitute equation (\ref{energyden}) into equation (\ref{menergy}),
we get the energy of the radiating charged particle in Einstein's universe
that is contained in a ``sphere'' of radius $r$
\be
E(r)=M(u)-\frac{4\pi Q^{2}(u)}{R}\cot\left(\frac{r}{R}\right)
\label{effmass}
\ee
which is also the energy (mass) of the gravitational field that a neutral particle
experiences at a finite distance $r$. Thus, the energy given
by equation (\ref{effmass}) is in addition called effective gravitational mass ($E=M_{eff}$)
of the spacetime under study. It is obvious that
the repulsive effect of the electric charge does not depend on its sign
and that the particle experiences a negative mass which acts repulsively when
\be
\frac{R}{\cot\left(\frac{r}{R}\right)}< \frac{4\pi Q^{2}(u)}{M(u)}
\label{condition}
\hspace{1ex}.
\ee
Additionally, if we replace equations (\ref{momden1}-\hspace{-0.1ex}
\ref{momden3}) into equation (\ref{mmomentum})
we get the momentum components which are given by
\bea
P_{1}=P_{2}=P_{3}=0\hspace{1ex}.
\eea

\par\noindent
Furthermore, we are interested in evaluating the power output,
i.e. the luminosity, for an observer at rest.
The power output is given by the expression
\be
L=-\frac{dM_{eff}}{du\,\,}
\label{luminosity}
\ee
and hence for the spacetime under study, it takes the form
\be
L=-M_{u}(u)+\frac{8\pi Q(u)Q_{u}(u)}{R}\cot\left(\frac{r}{R}\right)
\label{output}
\ee
where the subscript $u$ denotes the derivative with respect to the retarded null coordinate $u$.
One should not worry about the positivity of the power output due to the subtractive mass term
since as we have already mentioned, the mass function $M(u)$ is a nonincreasing function of the
retarded null coordinate $u$.

\par\noindent
As it was pointed out in the previous section several spacetimes are limiting cases of
gravitational background described by the metric (\ref{patel}). Therefore, it would be
interesting to derive the effective mass and the total output for these spacetimes
by taking the appropriate limits.
Some interesting spacetimes that are reduced as limiting cases of the metric
(\ref{patel}) are described below.

{\it (a) Einstein's universe}\\
As it was mentioned to the previous section, metric (\ref{patel}) becomes
the Einstein's universe when the mass and the charge of the particle are set equal to zero, i.e.
\be
ds^{2} = 2dudr +
du^{2} - R^{2}\sin^{2}\left(\frac{r}{R}\right)\left(d\theta^{2}+\sin^{2}\theta d\phi^{2}\right)
\hspace{1ex}.
\ee
It is easily seen from equation (\ref{effmass}) that the effective gravitational mass
of Einstein's universe is zero, i.e.
\be
M_{eff}^{Einstein}=0
\ee
which is exactly the same result that it was recently derived by Vargas \cite{vargas} who employed the
Einstein's and Landau-Lifshitz's energy-momentum complexes
in the framework of teleparallel gravity.
Additionally, the power output, as given by (\ref{output}), of the Einstein's universe is zero, i.e.
\be
L^{Einstein}=0
\hspace{1ex}.
\ee

{\it (b) Bonnor- Vaidya spacetime}\\
The Bonnor-Vaidya spacetime metric (\ref{bonnor})
which describes a radiating charged particle in flat spacetime, can be obtained from
the metric (\ref{patel}) when $R$ tends to infinity. It is evident from equation (\ref{effmass})
that the effective gravitational mass of Bonnor-Vaidya spacetime is given by
\be
M_{eff}^{B-V}=M(u)-\frac{4\pi Q^{2}(u)}{r}
\hspace{1ex}.
\ee
This is twice the effective mass computed by Chamorro and Virbhadra \cite{vir7} who
utilized the Einstein's and Landau-Lifshitz's energy-momentum complexes
\footnote{For a short discussion on this discrepancy (the anomalous factor 2) see \cite{katz}.}.
Additionally, the power output, as given by (\ref{output}), of the Bonnor-Vaidya spacetime is
\be
L^{B-V}=-M_{u}(u)+\frac{8\pi Q(u)Q_{u}(u)}{r}
\hspace{1ex}.
\ee

{\it (c) Vaidya spacetime}\\
When the electric charge of the particle is zero, i.e. $Q(u)=0$, and
$R$ tends to infinity, metric  (\ref{patel}) reduces to the Vaidya radiating-star
metric, i.e.
\be
ds^{2} = 2dudr + \left(1-\frac{2M(u)}{r}\right)du^{2} -
r^{2}\left(d\theta^{2}+\sin^{2}\theta d\phi^{2}\right)
\hspace{1ex}.
\ee
It is evident from equation (\ref{effmass})
that the effective gravitational mass of Vaidya spacetime is given by
\be
M_{eff}^{Vaidya}=M(u)
\hspace{1ex}.
\ee
This is exactly the same result computed by Lindquist, Schwartz and Misner \cite{misner} who
utilized the Landau-Lifshitz's energy-momentum complex.\\
The power output, as given by (\ref{output}), of the Vaidya spacetime takes the form
\be
L^{Vaidya}=-M_{u}(u)
\hspace{1ex}.
\ee

{\it (d) Reissner-Nordstr\"om black hole solution}\\
When the mass and the electric charge of the particle are constants, and
$R$ tends to infinity, metric  (\ref{patel}) reduces to the Reissner-Nordstr\"om black hole
metric, i.e.
\be
ds^{2} = 2dudr + \left(1-\frac{2M}{r}+\frac{4\pi Q^{2}}{r^{2}}\right)du^{2} -
r^{2}\left(d\theta^{2}+\sin^{2}\theta d\phi^{2}\right)
\hspace{1ex}.
\ee
It is clear from equation (\ref{effmass})
that the effective gravitational mass of Reissner-Nordstr\"om black hole is given by
\be
M_{eff}^{R-N}=M-\frac{4\pi Q^{2}}{r}
\hspace{1ex}.
\ee
This result agrees with the results computed by de la Cruz and Israel \cite{cruz} who used
junction conditions across thin shells, by Cohen and Gautreau \cite{cohen}
who implemented the Whittaker's theorem,  by Cohen and de Felice  \cite{felice} who
performed the evaluation using Komar's integral for energy, and by others \cite{rn}.
The power output, as given by (\ref{output}), of the Reissner-Nordstr\"om black hole is
\be
L^{R-N}=0
\hspace{1ex}.
\ee
Condition (\ref{condition}) that has to be fulfilled in order the particle to experience a
repulsive effect of gravity, for the case of Reissner-Nordstr\"om black hole, takes the form
\be
r < \frac{4\pi Q^{2}}{M}
\hspace{1ex}.
\ee
\section{Conclusions}
In this work, we explicitly calculate the energy and momentum density distributions associated
with a metric that describes a radiating charged particle in Einstein's universe.
Additionally, the effective gravitational mass, i.e. the total energy, and the power output,
i.e. the luminosity, of the specific gravitational background are explicitly evaluated. The
concept of effective gravitational mass is related to the repulsive effects of gravitation. A condition
is given which when satisfied a particle experiences a negative mass which acts repulsive.
The luminosity and consequently the power output must be positive. Although the mass term in the
expression for the power output is subtractive we have set the mass function
to be a nonincreasing function of the retarded null coordinate and therefore to guarantee the positivity of
power output. Since several interesting spacetimes are limiting cases of the gravitational
background under study, we have derived the effective mass and power output for these spacetimes
as limiting quantities. These results agree with the corresponding ones in the literature which were
computed by using different prescriptions and/or methods.
It is obvious that our results presented here provide evidence in support of Lessner statement for the
significance of M\"oller's prescription.
\section*{Acknowledgements}
This work has been supported by the European Research and Training Network
``EUROGRID-Discrete Random Geometries:
from Solid State Physics to Quantum Gravity" (HPRN-CT-1999-00161).

\end{document}